\documentclass[sigconf, natbib=true]{acmart}
\usepackage{subfigure} 
\usepackage[ruled]{algorithm2e}
\usepackage{algorithmic}
\usepackage{wasysym}
\usepackage{enumitem}
\usepackage{multirow}

\AtBeginDocument{%
  \providecommand\BibTeX{{%
    \normalfont B\kern-0.5em{\scshape i\kern-0.25em b}\kern-0.8em\TeX}}}

\setcopyright{acmlicensed}
\copyrightyear{2024}
\acmYear{2024}
\acmDOI{10.1145/3626772.3657738}

\setcopyright{acmlicensed}\acmConference[SIGIR '24]{Proceedings of the 47th
International ACM SIGIR Conference on Research and Development in
Information Retrieval}{July 14--18, 2024}{Washington, DC, USA}
\acmBooktitle{Proceedings of the 47th International ACM SIGIR Conference on
Research and Development in Information Retrieval (SIGIR '24), July 14--18,
2024, Washington, DC, USA}

\acmISBN{979-8-4007-0431-4/24/07}

\begin{document}

\title{Exploring the Individuality and Collectivity of Intents behind Interactions for Graph Collaborative Filtering}

% \affiliation{%

\author{Yi Zhang}
\affiliation{%
  \institution{Anhui University}
  \city{Hefei}
  \country{China}}
\email{zhangyi.ahu@gmail.com}

\author{Lei Sang}
\affiliation{%
 \institution{Anhui University}
  \city{Hefei}
  \country{China}
}
\email{sanglei@ahu.edu.cn}

\author{Yiwen Zhang}
\authornote{Yiwen Zhang is the corresponding author.}
\affiliation{%
  \institution{Anhui University}
  \city{Hefei}
  \country{China}
}
\email{zhangyiwen@ahu.edu.cn}

\renewcommand{\shortauthors}{Yi Zhang, Lei Sang, \& Yiwen Zhang}

\begin{abstract}
Intent modeling has attracted widespread attention in recommender systems. As the core motivation behind user selection of items, intent is crucial for elucidating recommendation results. The current mainstream modeling method is to abstract the intent into unknowable but learnable shared or non-shared parameters. Despite considerable progress, we argue that it still confronts the following challenges: firstly, these methods only capture the coarse-grained aspects of intent, ignoring the fact that user-item interactions will be affected by collective and individual factors (\textit{e.g.}, a user may choose a movie because of its high box office or because of his own unique preferences); secondly, modeling believable intent is severely hampered by implicit feedback, which is incredibly sparse and devoid of true semantics. To address these challenges, we propose a novel recommendation framework designated as Bilateral Intent-guided Graph Collaborative Filtering (BIGCF). Specifically, we take a closer look at user-item interactions from a causal perspective and put forth the concepts of individual intent—which signifies private preferences—and collective intent—which denotes overall awareness.
To counter the sparsity of implicit feedback, the feature distributions of users and items are encoded via a Gaussian-based graph generation strategy, and we implement the recommendation process through bilateral intent-guided graph reconstruction re-sampling. Finally, we propose graph contrastive regularization for both interaction and intent spaces to uniformize users, items, intents, and interactions in a self-supervised and non-augmented paradigm. Experimental results on three real-world datasets demonstrate the effectiveness of BIGCF compared with existing solutions.
\end{abstract}

\begin{CCSXML}
<ccs2012>
   <concept>
       <concept_id>10002951.10003317.10003347.10003350</concept_id>
       <concept_desc>Information systems~Recommender systems</concept_desc>
       <concept_significance>500</concept_significance>
       </concept>
 </ccs2012>
\end{CCSXML}

\ccsdesc[500]{Information systems~Recommender systems}

\keywords{Recommender System, Collaborative Filtering, Intent Modeling, Graph Neural Network, Self-Supervised Learning}



\maketitle

\section{Introduction}

\begin{figure*}
  \includegraphics[width=\textwidth]{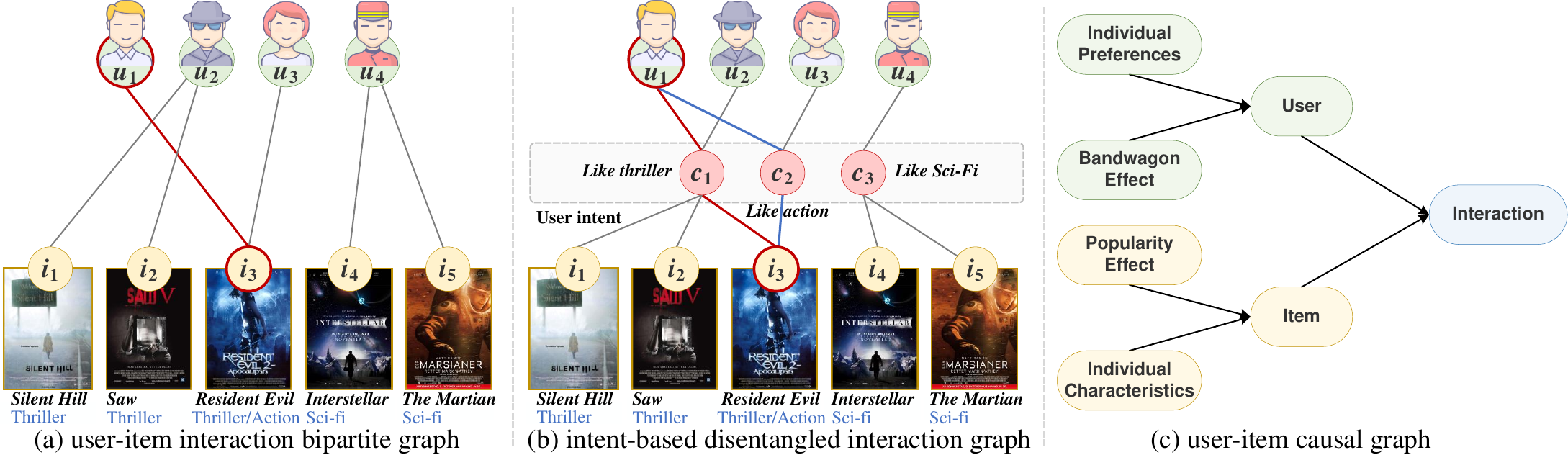}
  \caption{(a) A simplified user-item interaction bipartite graph in movie recommendation scenario; (b) disentangled interaction graph incorporating user intents, user $u_1$'s choice of film $i_3$ is influenced by intents $c_1$ and $c_2$, indicating that he likes horror films with action elements; (c) user-item causal graph considering collective and individual attributes.}

  \label{fig_motivation}
\end{figure*}

Personalized recommender systems \cite{ricci2011introduction} are the cornerstone of contemporary E-platforms for suggesting items to users that suit their needs. In terms of technology, recommender systems are built on the concept of collaborative filtering (CF) \cite{rendle2009bpr}, which aims to infer user preferences from historical interaction data \cite{he2017neural}. The main focus of current CF models is to develop high-quality embedding representations for both users and items. To achieve this goal, embedding modeling for CF models has been developed in a multifaceted way with a series of results, such as matrix factorization \cite{rendle2009bpr}, multilayer perceptrons \cite{he2017neural}, graph convolutional network (GCN) \cite{ying2018graph,wang2019neural,he2020lightgcn}, and graph contrastive learning (GCL) \cite{wu2021self, lin2022improving, ren2023disentangled}. Given that the user-item interactions naturally generate a bipartite graph (Fig. \ref{fig_motivation}(a)), the latest stage of the study is to investigate the provably motivating impacts of GCN \cite{kipf2016semi} on recommender systems. In concrete terms, LightGCN \cite{he2020lightgcn} has been the standard practice in graph-based recommender systems, but recent extensive studies strongly suggest that this is rapidly giving way to GCL \cite{you2020graph, wu2021self}.

% Given that the user-item interaction naturally generates a bipartite graph (Fig. \ref{fig_motivation}(a)), the latest stage of the study is to investigate the provably motivating impacts of Graph Convolutional Network (GCN) \cite{kipf2016semi} on recommender systems \cite{ying2018graph,wang2019neural,he2020lightgcn}. For example, NGCF and LightGCN attempt to encode higher-order representations of users and items on interaction graph. LightGCN has been the traditional practice in recommender system, but recent extensive studies strongly suggest that this is rapidly giving way to the Graph Contrastive Learning (GCL). For instance, SGL constructs view copies of the interaction graph by random masking, and HCCF constructs semantic representations at the interaction and hypergraph levels, respectively, and supplies extra supervised signals for the recommendation task through the contrastive process.

Despite the impressive results achieved by these methods, we argue that these attempts fall short of providing a more detailed explanation of user behavior. In contrast to the unintentional actions of animals, user behavior is constantly motivated by multiple factors and is susceptible to social influences \cite{wu2019neural}. In a nutshell, there are real \textit{intents} hidden behind the user's choice of items \cite{chang2023latent}. Intent for recommendation has been investigated in some groundbreaking studies. For example, DGCF \cite{wang2020disentangled} and KGIN \cite{wang2021learning} aim to learn multiple disentangled intent representations for users, whereas DCCF \cite{ren2023disentangled} also takes item-side intent modeling into account. We note that these efforts provide interpretability to the recommendation results while maintaining competitive performance, but do not fully reflect the user's individual preferences.

For a deeper understanding of the interactions, we show a refined disentangled interaction graph, as shown in Fig. \ref{fig_motivation}(b), and user behavior can be divided into two aspects. On the one hand, a user's behavior is not isolated but influenced by others \cite{wu2019neural}, which is also an essential requirement of collaborative filtering \cite{he2017neural}. Based on this fact, the set of intents $\{c_1,c_2,c_3\}$ in Fig. \ref{fig_motivation}(b) is not unique to user $u_1$, but is shared by all users. For example, the fact that a group of users $\{u_1,u_2\}$ like movies with horror themes suggests that the users' behavior come from a collective intent $c_1$. This is referred to as the \textit{Bandwagon Effect} in social psychology terminology \cite{knyazev2022bandwagon}. On the other hand, user behavior also exhibits individualism \cite{chen2017attentive}. As illustrated in Fig. \ref{fig_motivation}(b), not only does user $u_1$ enjoy horror movies, but he also enjoys horror movies that have action elements. In fact, the individual preference of user $u_1$ represents an intent set $\{c_1,c_2\}$.

Considering that items are not subjective, there are barriers to defining the intent of an item. Motivated by \cite{ma2019learning, ren2023disentangled}, we can perceive item intent as an attribute, meaning the reason the user selects it. Similar to users, items can likewise be portrayed in two aspects: the popularity effect and individual characteristics. In general, the \textit{Popularity Effect} reflects the affinity of items \cite{wei2021model}, which corresponds to the bandwagon effect on the user side. The greater the item's popularity within the user community, the higher the likelihood of it being selected. And the more substantial factors influencing users' decisions to pick an item are its individual characteristics. For example, movie $i_1$ has theme \textit{thriller}, while movie $i_3$ has themes \textit{thriller} and \textit{action} in Fig. \ref{fig_motivation}(b).

Based on the aforementioned study, the interactions between users and items are influenced by both from collective and individual, which can be abstracted into the fine-grained causal graph shown in Fig. \ref{fig_motivation}(c). To achieve accurate recommendations with intents, it is essential to consider these factors at a finer granularity. However, it will face the following two challenges:
\begin{itemize}[leftmargin=*]
\item[$\bullet$] How to consider both collective and individual factors for users (items) and realize adaptive trade-offs?
\item[$\bullet$] How to model user (item) intents using only sparse and semantic-free interaction data?
\end{itemize}

To tackle the above challenges, we propose a novel  \textbf{B}ilateral   \textbf{I}ntent-guided \textbf{G}raph  \textbf{C}ollaborative \textbf{F}iltering (\textbf{BIGCF}) for implicit feedback-based recommender system. Considering the sparsity of implicit feedback, we convert the graph recommendation problem into a graph generation task and encode the feature distributions of users and items via iterative GCNs. Such feature distributions are regarded as user preferences and item characteristics, respectively. More further, the bandwagon and popularity effects are abstracted as the \textbf{Collective Intents} on the user and item sides, respectively. Subsequently, we learn disentangled \textbf{Individual Intents} for users and items utilizing collective intents with user preference and item characteristic distributions. Definitionally, the collective intents are learnable parameters shared by all users (items), whereas the individual intents are linear combinations of user preferences (item characteristics) with collective intents. Finally, inspired by the recent motivating effects of GCL in recommender systems \cite{you2020graph, wu2021self}, we propose augmentation-agnostic \textbf{Graph Contrastive Regularization} in both interaction and intent spaces, respectively, to self-supervisedly attain uniformity and alignment for all nodes on interaction graph. Overall, BIGCF is an extremely simplified framework for personalized recommender systems that takes into account the individual and collective factors of user-item interactions and enhances recommendation performance through bilateral intent modeling. The major contributions of this paper are summarized as follows:
\begin{itemize}[leftmargin=*]
\item[$\bullet$] We decompose the motivations behind user-item interactions into collective and individual factors, and propose the recommendation framework BIGCF, which focuses on exploring the individuality and collectivity of intents on the interaction graph.
\item[$\bullet$] We further propose graph contrastive regularization in both interaction and feature spaces, which regulates the uniformity of the whole feature spaces in an augmentation-agnostic manner and achieves collaborative cross-optimization in dual spaces.
\item[$\bullet$] We conduct extensive experiments on three public datasets and show that BIGCF not only significantly improves recommendation performance but can also effectively explore users' true preferences.
\end{itemize}

\begin{figure*}[t]
\setlength{\abovecaptionskip}{0.1cm}
\setlength{\belowcaptionskip}{0.1cm} 
\centering
\includegraphics[width=\linewidth]{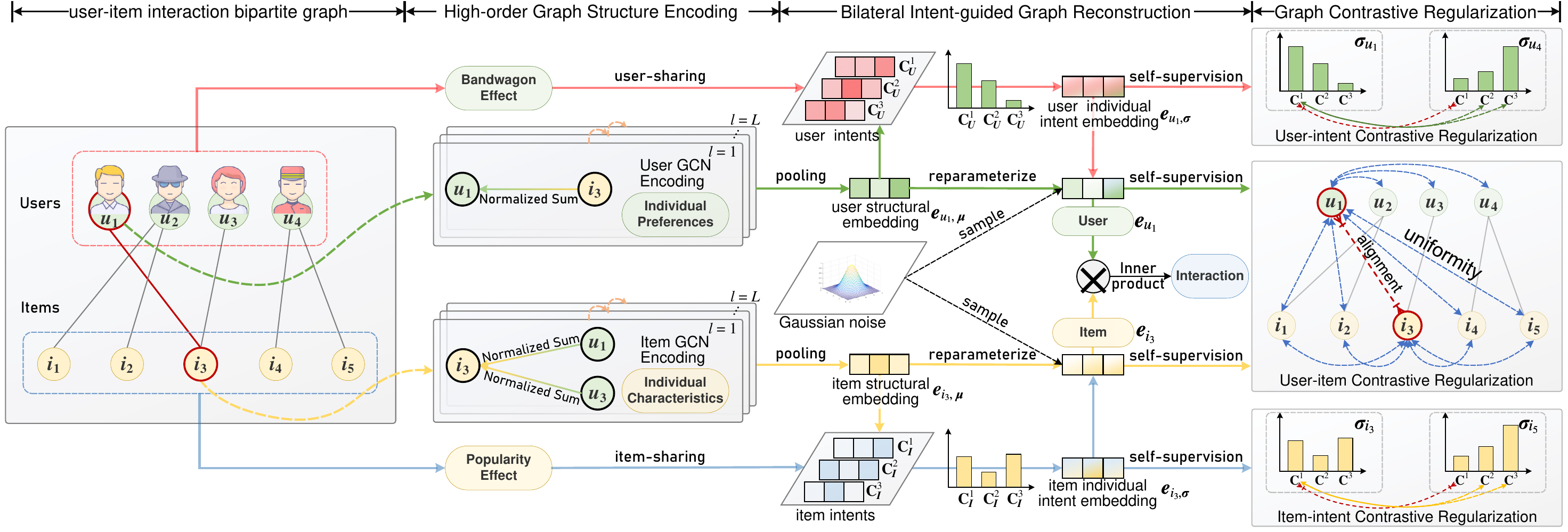}
\caption{The complete framework of the proposed BIGCF. BIGCF consists of a high-order graph structure encoding, a bi-directional intent-directed graph reconstruction and a graph contrastive regularization process in dual space.}
\label{fig_model}
\end{figure*}

\section{METHODOLOGY}
\subsection{Problem Formulation}
A typical recommendation scenario includes a set of $M$ users $\mathcal U=\{u_1, u_2, \dots, u_M\}$ and a set of $N$ items $\mathcal I=\{i_1, i_2, \dots, i_N\}$. Furthermore, a user-item interaction matrix $\mathbf R \in \mathbb {R}^{M\times N}$ is given according to the historical user-item interactions. Based on previous works \cite{wang2019neural,he2020lightgcn}, interactions can be abstracted to a bipartite graph structure $\mathcal G=<\mathcal V =\{\mathcal U, \mathcal V\}, \mathcal E>$, where $\mathcal E_{ui}=\mathcal E_{iu}=R_{ui}$. Therefore, the recommendation task can also be considered as a bilateral graph generation problem \cite{truong2021bilateral}, \textit{i.e.}, predicting the probability of the existence of edges on the user-item interaction graph $\mathcal G$: 
\begin{align}
\begin{aligned}
\mathbb P(\hat {\mathbf R}|\mathbf E^{(0)},\mathbf R) &  =\prod_{u \in \mathcal U}\prod_{i\in\mathcal I}\mathbb P(\hat R_{ui}|\mathbf e_u,\mathbf e_i),
\end{aligned}
\end{align}
where $\mathbf E^{(0)}=\{\mathbf E^{(0)}_{\mathcal U},\mathbf E^{(0)}_{\mathcal I}\}$ is the initial interaction embedding table. $\mathbf e_u$ and $\mathbf e_i$ are the embeddings of user $u$ and item $i$ via encoding. The widely adopted graph recommendation paradigm is to model user preferences leveraging interactions, a process that relies solely on the interaction graph $\mathcal G$ \cite{he2020lightgcn, zhang2024nie}. However, user behavior is often driven by a variety of intents, which are complex and intermingled, with some users gravitating towards horror movies while others have alternatives. At the probabilistic level, the interaction probability requires additional consideration of the effect of intents $\mathbf C$:
\begin{equation}
\begin{aligned}
\label{intent_p}
\mathbb P(\hat {\mathbf R}|\mathbf E^{(0)},\mathbf R,\mathbf C)=\prod_{u \in \mathcal U}\prod_{i\in\mathcal I}\mathbb P(\hat R_{ui}|\mathbf e_u,\mathbf e_i)\times\\
\sum_{k \in \mathcal K}\mathbb P(\mathbf e_u|\mathbf E^{(0)},\mathbf R,\mathbf C_{\mathcal U}^k)\mathbb P(\mathbf e_i|\mathbf E^{(0)},\mathbf R,\mathbf C_{\mathcal I}^k),
\end{aligned}
\end{equation}
where $\mathbf C=\{\mathbf C_{\mathcal U}^k,\mathbf C_{\mathcal I}^k|k \in \mathcal K\}$ is the collective intent table for all user and item nodes, of which the number is $|\mathcal K|$. Intent $\mathbf C$ is shared by all users and items as a way to adequately model the collectivity of user preferences and item characteristics. As described in the Introduction section, user intent $\mathbf C_{\mathcal U}$ represents the bandwagon effect with group volition, indicating the common motivation for choosing items, while item intent $\mathbf C_{\mathcal I}$ means the popularity effect, indicating the common reasons for why the item is selected. Fig. \ref{fig_model} presents the framework of BIGCF, which is a refined version of the user-item causal graph given in Fig. \ref{fig_motivation}(c).

\subsection{BIGCF}
To fully understand user (item) intents and make recommendations, we need to model the interaction and intent embeddings for user $u$ and item $i$, and construct the final embedding representations $\mathbf e_u$ and $\mathbf e_i$ that are used for recommendations $\mathbb P(\hat R_{ui}|\mathbf e_u,\mathbf e_i)$. Considering the sparsity and semantic-free nature of implicit feedback, we represent the embeddings $\mathbf e_u$ and $\mathbf e_i$ as Gaussian distributions with the support of variational inference \cite{kingma2013auto, liang2018variational}:
\begin{equation}
\begin{aligned}
\label{vae}
\mathbb P(\mathbf e_u|\mathbf E^{(0)},\mathbf R,\mathbf C_{\mathcal U})&\sim\mathcal N\left (\mathbf e_u|\boldsymbol{\mu}_u, \text{diag}[\boldsymbol{\sigma}^2_u]  \right ), \\
\mathbb P(\mathbf e_i|\mathbf E^{(0)},\mathbf R,\mathbf C_{\mathcal I})&\sim\mathcal N\left (\mathbf e_i|\boldsymbol{\mu_i}, \text{diag}[\boldsymbol{\sigma}^2_i]  \right ) ,
\end{aligned}
\end{equation}
where $\boldsymbol{\mu}_u$ and $\boldsymbol{\sigma}^2_u$ are the approximate mean and variance of the isotropic Gaussian Distribution of node $u$, respectively. 

\subsubsection{\textbf{High-order Graph Structure Encoding}}
Many studies have demonstrated that the GCN-based embedding encoding process \cite{kipf2016semi} can fully utilize the rich collaborative relationships on the user-item interaction graph $\mathcal G$ \cite{he2020lightgcn, wu2021self}. Considering that BIGCF is a model-agnostic recommendation framework, we empirically employ lightweight GCN \cite{he2020lightgcn} to encode high-order collaborative relationships:
\begin{equation}
\label{gcn_mu}
\mathbf e_{u,\boldsymbol{\mu}}^{(l)}=\sum_{i\in \mathcal S_u}\frac{1}{\sqrt{|\mathcal S_u||\mathcal S_i|}}\mathbf e_{i,\boldsymbol{\mu}}^{(l-1)},\quad \mathbf e_{i,\boldsymbol{\mu}}^{(l)}\sum_{u\in \mathcal S_i}\frac{1}{\sqrt{|\mathcal S_u||\mathcal S_i|}}\mathbf e_{u,\boldsymbol{\mu}}^{(l-1)},
\end{equation}
where $\mathbf e_{u,\boldsymbol{\mu}}^{(l)}$ and $\mathbf e_{i,\boldsymbol{\mu}}^{(l)}$ are $l$-th layer ($l \in [1, L]$) user and item structural embeddings, respectively (we define $\mathbf e_{u,\boldsymbol{\mu}}^{(0)}=\mathbf e_u^{(0)}$ and $\mathbf e_{i,\boldsymbol{\mu}}^{(l)}=\mathbf e_i^{(0)}$), and $\mathcal S_u$ and $\mathcal S_i$ are the one-order receptive field of user $u$ and item $i$, respectively. From a macroscopic view, the process is a cumulative multiplication of the graph Laplace matrix: $\mathbf E_{\boldsymbol{\mu}}^{(l)} = \left (\mathbf D^{-0.5}\mathbf A \mathbf D^{-0.5}\right)^l \mathbf E^{(0)}$, where $\mathbf D$ is the diagonal degree matrix of $\mathbf A$, and $\mathbf A \in \mathbb R^{(M+N)\times (M+N)} $ is the adjacent matrix of $\mathcal G$. After $L$ layers of propagation, we empirically construct the final structural embeddings by sum pooling for speed: $\mathbf e_{u,\boldsymbol{\mu}}={\textstyle \sum_{l=0}^{L}\mathbf e_{u,\boldsymbol{\mu}}^{(l)}}$ and $\mathbf e_{i,\boldsymbol{\mu}}={\textstyle \sum_{l=0}^{L}\mathbf e_{i,\boldsymbol{\mu}}^{(l)}}$. Considering the variability of neighbor contributions to the central node at each order on the interaction graph $\mathcal G$, the computation of $\mathbf E_{\boldsymbol{\mu}} = \{\mathbf E_{\mathcal U, \boldsymbol{\mu}},\mathbf E_{\mathcal I, \boldsymbol{\mu}} \}$ is included in each layer to capture user preferences and item characteristics with different structural semantics.

\subsubsection{\textbf{Bilateral Intent-guided Graph Reconstruction}}
We now switch perspectives to investigate the correspondence between intents and interactions. Given the unique preferences of the user $u$ and the unique characteristics of the item $i$, we need to additionally consider the adaptability between intents and interactions. Specifically, we first measure the correlation scores between the structural embeddings $\{\mathbf e_{u,\boldsymbol{\mu}}, \mathbf e_{i,\boldsymbol{\mu}}\}$ and the collective intents $\{\mathbf C_{\mathcal U},\mathbf C_{\mathcal I}\}$:
\begin{equation}
\begin{aligned}
\label{softmax_mu}
\mathbb P(\mathbf C_{\mathcal U}^k|\mathbf e_{u,\boldsymbol{\mu}})&=\text{exp}\left ({\mathbf e_{u,\boldsymbol{\mu}}}^{\top}\mathbf C_{\mathcal U}^k /\kappa\right )/{\textstyle \sum_{k'}^{\mathcal K}}\text{exp}\left ({\mathbf e_{u,\boldsymbol{\mu}}}^{\top}\mathbf C_{\mathcal U}^{k'}/\kappa\right),\\
\mathbb P(\mathbf C_{\mathcal I}^k|\mathbf e_{i,\boldsymbol{\mu}})&=\text{exp}\left ({\mathbf e_{i,\boldsymbol{\mu}}}^{\top}\mathbf C_{\mathcal I}^k /\kappa\right ) /{\textstyle \sum_{k'}^{\mathcal K}}\text{exp}\left ({\mathbf e_{i,\boldsymbol{\mu}}}^{\top}\mathbf C_{\mathcal I}^{k'}/\kappa\right) ,
\end{aligned}
\end{equation}
where $\kappa \in [0,1]$ is a temperature coefficient used to adjust the predicted distribution. In practice, the collective intent $\{\mathbf C_{\mathcal U},\mathbf C_{\mathcal I}\}$ is utilized to compute the affinity between the interaction and each intent $k \in \mathcal K$, reflecting the individuality and enabling deeper integration of disentangled intents with the graph structure:
\begin{equation}
\label{gcn_sigma}
\mathbf e_{u,\boldsymbol{\sigma}}=\sum_{k \in \mathcal K} \mathbb P(\mathbf C_{\mathcal U}^k|\mathbf e_{u,\boldsymbol{\mu}})\cdot \mathbf C_{\mathcal U}^k, \quad
\mathbf e_{i,\boldsymbol{\sigma}}=\sum_{k \in \mathcal K} \mathbb P(\mathbf C_{\mathcal I}^k|\mathbf e_{i,\boldsymbol{\mu}})\cdot \mathbf C_{\mathcal I}^k,
\end{equation}
where $\mathbf e_{u,\boldsymbol{\sigma}}$ and $\mathbf e_{i,\boldsymbol{\sigma}}$ are the individual intent embeddings of user $u$ and item $i$, respectively. In order to obtain the approximate posterior that can fit the user's preferences, we regard the encoded structural embeddings and the individual intent embeddings as the variational parameters for distributions defined in Eq. \ref{vae}. Considering the additivity of the Gaussian distributions \cite{ho2020denoising}, varying intents are seen as a combination of approximate Gaussians to adequately fit the individualized profile of user $u$ (or item $i$):
\begin{equation}
\begin{aligned}
\label{multi_gaussian}
\mathbb P(\mathbf e_u|\mathbf E^{(0)},\mathbf R,\mathbf C_{\mathcal U})&=\sum_{k \in \mathcal K}\mathbb P(\mathbf e_u|\mathbf E^{(0)},\mathbf R,\mathbf C_{\mathcal U}^k)\\&\sim \mathcal N\left (\mathbf e_u|\mathbf e_{u,\boldsymbol{\mu}}, \sum_{k\in \mathcal K}\mathbb P(\mathbf C_{\mathcal U}^k|\mathbf e_{u,\boldsymbol{\mu}})\cdot \mathbf C_{\mathcal U}^k  \right )\\&\sim\mathcal N\left (\mathbf e_u|\mathbf e_{u,\boldsymbol{\mu}}, \mathbf e_{u,\boldsymbol{\sigma}}   \right ).
\end{aligned}
\end{equation}
The item side has a similar definition. The mean $\mathbf e_{u,\boldsymbol{\mu}}$ represents the individual preferences for user $u$, which are independent of intents, while the variance $\mathbf e_{u,\boldsymbol{\sigma}}$ is a linear combination of distributions generated by multiple intents, which provides the variance with a variety of implications. For example, the user $u_1$ in Fig. \ref{fig_motivation}(b) will contain more correlation scores from intents $c_1$ and $c_2$. And the intent $c_1$ should have the highest score because the primary premise for choosing this movie is the horror theme. Thus we can assume that $\mathbf e_{u_1,\boldsymbol{\sigma}} = 0.6c_1 + 0.3c_2 + 0.1c_3$.

Given a user $u$ and an item $i$, the mean and variance embeddings of the approximate posterior can be used to sample $\mathbf e_u \sim \mathcal N(\boldsymbol{\mu}_u, \boldsymbol{\sigma}_u^{2})$ and $\mathbf e_i \sim \mathcal N(\boldsymbol{\mu}_i, \boldsymbol{\sigma}_i^{2})$. Considering the non-differentiable nature of the sampling process, reparameterization trick \cite{kingma2013auto} is used here to cleverly circumvent the back-propagation problem:
\begin{equation}
\begin{aligned}
\label{reparameterization}
\mathbb P(\mathbf e_u|\mathbf E^{(0)},\mathbf R,\mathbf C_{\mathcal U})&\sim \mathbf e_{u,\boldsymbol{\mu}} +  {\mathbf e_{u,\boldsymbol{\sigma}}}\odot \boldsymbol\epsilon_u, \\ \mathbb P(\mathbf e_i|\mathbf E^{(0)},\mathbf R,\mathbf C_{\mathcal I})&\sim \mathbf e_{i,\boldsymbol{\mu}} + \mathbf e_{i,\boldsymbol{\sigma}}\odot \boldsymbol\epsilon_i,
\end{aligned}
\end{equation}
where $\boldsymbol\epsilon_u$ and $\boldsymbol\epsilon_i$ are auxiliary noise variables sampled from $\mathcal N(\mathbf 0, \mathbf I)$, and $\odot$ denotes an element-wise product. We emphasize the advantages of introducing reparameterization in the following ways: (1) $\boldsymbol\epsilon_u$ and $\boldsymbol\epsilon_i$ achieve fusion in an adaptive manner, thus avoiding tedious manual adjustments; (2) Eq. \ref{reparameterization} does not depend on any additional process, such as attention mechanisms. Combining Eqs. \ref{gcn_mu}, \ref{softmax_mu}, \ref{gcn_sigma}, \ref{multi_gaussian}, and \ref{reparameterization}, we can obtain:
\begin{equation}
\begin{aligned}
\label{final_gcn}
\mathbf e_u & =  \underbrace{\sum_{l=1}^{L}\sum_{i\in \mathcal S_u}\frac{1}{\sqrt{|\mathcal S_u||\mathcal S_i|}}\mathbf e_i^{(l-1)}}_{\text{structural information:}\mathbf e_{u,\boldsymbol{\mu}} } +(\underbrace{\sum_{k \in \mathcal K} \mathbb P(\mathbf C_{\mathcal U}^k|\mathbf e_{u,\boldsymbol{\mu}})\cdot \mathbf C_{\mathcal U}^k}_{\text{user intent:}\mathbf e_{u,\boldsymbol{\sigma}}  })\odot \boldsymbol\epsilon _u,\\
\mathbf e_i &= \underbrace{\sum_{l=1}^{L}\sum_{u\in \mathcal S_i}\frac{1}{\sqrt{|\mathcal S_u||\mathcal S_i|}}\mathbf e_u^{(l-1)}}_{\text{structural information:}\mathbf e_{i,\boldsymbol{\mu}} } +(\underbrace{\sum_{k \in \mathcal K} \mathbb P(\mathbf C_{\mathcal I}^k|\mathbf e_{i,\boldsymbol{\mu}})\cdot \mathbf C_{\mathcal I}^k}_{\text{item intent:}\mathbf e_{i,\boldsymbol{\sigma}}  })\odot \boldsymbol\epsilon _i.
\end{aligned}
\end{equation}
Eq. \ref{final_gcn} further highlights the fusion process of structural information and intents based on the reparameterization process. It is important to note that the intent $\mathbf C_{\mathcal U}$ ($\mathbf C_{\mathcal I}$) is common to all users (items) and represents the collectivity of users (items), however the variance vector $\mathbf e_{u,\boldsymbol{\sigma}}$ ($\mathbf e_{i,\boldsymbol{\sigma}}$) is unique and represents the individuality of user $u$ (item $i$) on the interaction graph $\mathcal G$.

Based on the fused embeddings, we compute the probability scores of the existence of interaction edges $R_{ui}$ via inner product:
\begin{equation}
\label{score}
\mathbb P(\hat R_{ui}|\mathbf e_u,\mathbf e_i)=\text{Sigmoid}(\mathbf e_u^\top \mathbf e_i).
\end{equation}

\subsubsection{\textbf{Graph Contrastive Regularization in Dual Spaces}}
There is evidence from a number of recent research that self-supervised learning (SSL) provides extra supervised signals to assist ease the data sparsity problem, and is positively motivating for recommendation tasks \cite{wu2021self}. Contrastive learning \cite{chen2020simple}, which builds a pair of views of anchor nodes via data augmentation and maximizes the mutual information between them, is generally the most popular strategy for self-supervised learning. Given arbitrary node $a$, the widely used contrastive loss infoNCE \cite{chen2020simple} is defined as follows:
\begin{equation}
\label{infoNCE}
\mathcal I(\mathbf a', \mathbf a'')=-\text{log}\frac{\text{exp}\left (\text{cos}\left ( {\mathbf a',\mathbf a''}\right)/\tau\right)}{ {\textstyle \sum_{b \in \mathcal B}{\text{exp}(\text{cos}\left ( {\mathbf a',\mathbf b}\right)/\tau)}}},
\end{equation}
where $\mathbf a'$ and $\mathbf a''$ are two augmented embeddings of $a$, $\text{cos}(\cdot,\cdot)$ means the cosine similarity, and $\tau$ is a predefined temperature coefficient. Note that the other nodes $\{b \in \mathcal B\}$ in a mini-batch are considered as negative samples of node $a$ in Eq. \ref{infoNCE} for simplicity. Most of existing recommendation paradigms adopt various types of data augmentation to perform SSL task, such as structural \cite{wu2021self, yang2022knowledge}, feature \cite{lin2022improving}, or semantic augmentation \cite{ren2023disentangled}. Nevertheless, these augmentation techniques not only add significantly to the training cost but also have a high propensity to distort the semantic information of input data. For example, removing key node structures may destroy the original interaction graph $\mathcal G$.

Noting that the core of infoNCE lies in uniformizing the feature space \cite{wang2020understanding, wang2021understanding}, we regard the contrastive process $\mathcal I(\mathbf a, \mathbf b)$ as a regularization of all nodes $\mathcal V$ on the interaction graph $\mathcal G$. Specifically, regularization in the interaction space consists of two parts:
\begin{itemize}[leftmargin=*]
\item[$\bullet$] \textbf{Interaction Regularization (IR) $\mathcal I(\mathbf e_u, \mathbf e_i)$}: Considering that the essence of recommendation lies in quantifying the user-item similarity, it is necessary to ensure a closer distance between the user $u$ and the positive interaction $i$, and it is also needed to keep the user $u$ as far away as possible from the negatives $\{j\in \mathcal B/i\}$.
\item[$\bullet$] \textbf{Homogeneous Node Regularization (HNR)} $\mathcal I(\mathbf e_u, \mathbf e_u)$ and $\mathcal I(\mathbf e_i, \mathbf e_i)$:  The Matthew effect caused by the popularity bias \cite{wei2021model} also needs to be taken into account to prevent the learned embeddings from concentrating close to one direction, which would result in dimensional collapse \cite{jing2021understanding}. We keep the nodes themselves (\textit{e.g.,} $u$) as positive samples while keeping different nodes (\textit{e.g.,} $\{v\in \mathcal B/u\}$) far away from each other.

\end{itemize}
Based on the above analysis, we propose the graph contrastive regularization process in the interaction space:
\begin{equation}
\begin{aligned}
\label{graphReg_1}
\mathcal L_{\text{inter}}=\sum_{<u,i> \in \mathcal B}\left(\mathcal I(\mathbf e_u, \mathbf e_i) + \mathcal I(\mathbf e_u, \mathbf e_u) + \mathcal I(\mathbf e_i, \mathbf e_i)\right),\\
\end{aligned}
\end{equation}
where $\mathbf e_u$ and $\mathbf e_i$ are the user and item embeddings obtained via Eq. \ref{final_gcn}. We emphasise the superiority of Eq. \ref{graphReg_1} in two respects. Firstly, the loss $\mathcal L_{\text{inter}}$ does not depend on any form of data augmentation, which not only significantly reduces the time complexity but also ensures the integrity of the structural information. Secondly, Eq. \ref{graphReg_1} guarantees alignment between positive interaction pairs while keeping the entire interaction feature space uniform, whereas traditional CL strategies do not take into account the alignment relationships between users and items.

\begin{figure}[t]
\setlength{\abovecaptionskip}{0.2cm}
\setlength{\belowcaptionskip}{0.2cm} 
\centering
\includegraphics[width=\linewidth]{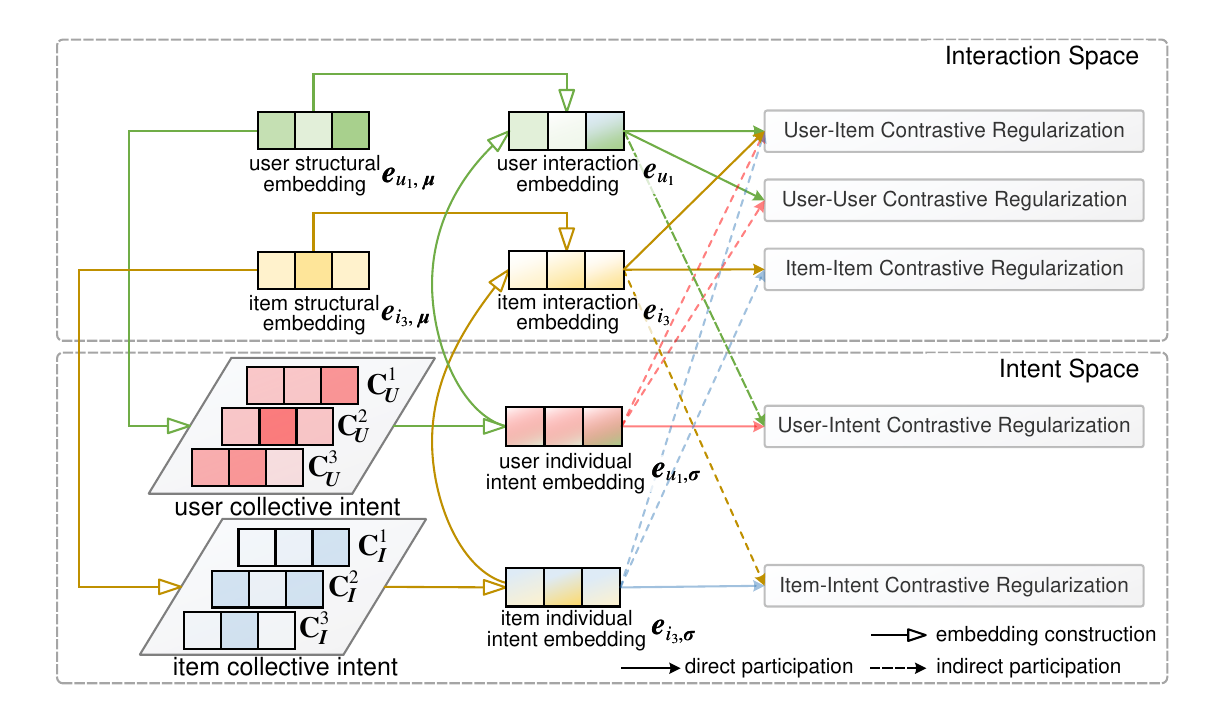}
\caption{The computational process and information flow of graph contrastive regularization in interaction space and intent space. The solid line indicates that the input is directly involved in the computation, while the dashed line indicates that this input is indirectly involved in the operation. }
\label{fig_graph_reg}
\end{figure}

\begin{itemize}[leftmargin=*]
\item[$\bullet$] \textbf{Bilateral Intent Regularization (BIR)} $\mathcal I(\mathbf e_{u, \boldsymbol{\sigma}}, \mathbf e_{u, \boldsymbol{\sigma}})$ and $\mathcal I$ $(\mathbf e_{i, \boldsymbol{\sigma}}, \mathbf e_{i, \boldsymbol{\sigma}})$: In the intent space, we obtain the individual intent embeddings $\{\mathbf e_{u, \boldsymbol{\sigma}},\mathbf e_{i, \boldsymbol{\sigma}}\}$ of user $u$ and item $i$ through the collective intents $\{\mathbf C_{\mathcal U},  \mathbf C_{\mathcal I}\}$.
Considering that intent embeddings $\{\mathbf e_{u, \boldsymbol{\sigma}},\mathbf e_{i, \boldsymbol{\sigma}}\}$ indicate different motivations (reasons) for user $u$ (item $i$) to select (be selected) items (by users), if an intent $k$ can be represented by other intents $\{k' \in \mathcal K / k\}$, then such an intent $k$ may be redundant.
\end{itemize}
 Based on this, we further model disentangled intent representations via graph contrastive regularization process in the intent space:
\begin{equation}
\label{graphReg_2}
\mathcal L_{\text{intent}}=\sum_{u \in \mathcal B}\mathcal I(\mathbf e_{u, \boldsymbol{\sigma}}, \mathbf e_{u, \boldsymbol{\sigma}}) +  \sum_{i\in \mathcal B}\mathcal I(\mathbf e_{i, \boldsymbol{\sigma}}, \mathbf e_{i, \boldsymbol{\sigma}}).
\end{equation}
 Similar to Eq. \ref{graphReg_1}, $\mathcal L_{\text{intent}}$ also does not rely on any form of data augmentation. The process of graph contrastive regularization in both interaction and intent spaces is shown in Fig. \ref{fig_graph_reg}. Notice that the computational chains of final interaction embeddings $\{\mathbf e_{u},\mathbf e_{i}\}$ and individual intent embeddings $\{\mathbf e_{u, \boldsymbol{\sigma}},\mathbf e_{i, \boldsymbol{\sigma}}\}$ contain the collective intent $\{\mathbf C_{\mathcal U},  \mathbf C_{\mathcal I}\}$ and the structural embeddings $\{\mathbf e_{u, \boldsymbol{\mu}},\mathbf e_{i, \boldsymbol{\mu}}\}$, which allows the proposed graph contrastive regularization process to achieve cross-domain collaboration in dual spaces. We will further analyze in Subsection \ref{analysis}.

\subsection{Multi-task Joint Training}

Recommendation tasks based on graph structures can be viewed as graph generation problems, \textit{i.e.}, predicting the likelihood of the existence of interaction edges \cite{zhang2023revisiting}. Given the interaction graph $\mathcal G$ and interaction matrix $\mathbf R$, we assume that arbitrary data point comes from the generation process $\mathbf R \sim \mathbb P(\hat{\mathbf R}|\mathbf E)$. In BIGCF, we learn the mean and variance embeddings of the approximate posterior for user and item, \textit{i.e.}, $\mathbf E\sim \mathbb P(\mathbf E|\Theta,\mathbf R)$.  Then the graph generation task can be optimized by the Evidence Lower Bound (ELBO) \cite{kingma2013auto}:
\begin{equation}
\label{ELBO}
  \mathcal L_{\text{ELBO}}=\mathbb E_{\mathbf E\sim \mathbb P(\mathbf E|\Theta,\mathbf R)}[\text{log}(\mathbb P(\hat{\mathbf R}|\mathbf E)]-\text{KL}(\mathbb P(\mathbf E|\Theta,\mathbf R)||\mathbb P(\mathbf E)),
\end{equation}
where $\Theta=\{\mathbf E^{(0)}, \mathbf C\}$ is the set of model parameters. The first term is the reconstruction error between the input interaction matrix $\mathbf R$ and the reconstructed interaction matrix $\hat{\mathbf R}$, and in line with existing methods \cite{wang2019neural, he2020lightgcn, wu2021self}, we empirically employ the pair-wise loss \cite{rendle2009bpr} to guide the reconstruction process on the interaction graph $\mathcal G$:

\begin{equation}
\label{BPR}
\mathbb E_{\mathbb P(\mathbf E|\Theta,\mathbf R)}[\text{log}(\mathbb P(\hat{\mathbf R}|\mathbf E)]=\sum_{<u,i,j> \in \mathcal B}-\text{log}\delta(\mathbb P(\hat R_{ui}) - \mathbb P(\hat R_{uj}))
\end{equation}
where $\delta$ is the Sigmoid activation function, $\mathbb P(\hat{R}_{ui})$ and $\mathbb P(\hat{R}_{uj})$ are short for $\mathbb P(\hat{R}_{ui}|\mathbf e_u,\mathbf e_i)$ and $\mathbb P(\hat{R}_{uj}|\mathbf e_u,\mathbf e_j)$, respectively, and $i$ and $j$ are positive and negative samples of user $u$, respectively. The second term in Eq. \ref{ELBO} is the KL divergence between the posterior distribution $\mathbb P(\mathbf E|\Theta,\mathbf R)$ and the prior $\mathbb P(\mathbf E)$. For simplicity, the prior $\mathbb P(\mathbf E)$ is set to a standard Gaussian $\mathcal N(\mathbf 0, \mathbf I)$ \cite{liang2018variational}.

Finally, to integrate recommendation task and intent modeling, we employ a multi-task joint training strategy to optimise the main recommendation task and the graph contrastive regularization in both interaction and intent spaces. The complete optimization objective of BIGCF is defined as follows:

\begin{equation}
\label{loss}
\mathcal L_{\text{BIGCF}} = \mathcal L_{\text{ELBO}} + \lambda_1 (\mathcal L_{\text{inter}} + \mathcal L_{\text{intent}}) + \lambda_2 \left \|\Theta  \right \|_2^2
\end{equation}
where $\lambda_1$ and $\lambda_2$ are adjustable weights, and $\Theta=\{\mathbf E^{(0)}, \mathbf C\}$ are trainable model parameters.

\subsection{Model Analysis}
% \subsubsection{\textbf{Space Complexity}}
% The model parameters of BIGCF include initial user and item embeddings $\mathbf E^{(0)}=\{\mathbf E^{(0)}_{\mathcal U} \in \mathbb R^{M \times d}, \mathbf E^{(0)}_{\mathcal I} \in \mathbb R^{N \times d} \} \in \mathbb R^{(M+N)\times d}$, as well as additionally defined commonality intent embedding tables $\mathbf C=\{\mathbf C_{\mathcal U} \in \mathbb R^{|\mathcal K| \times d}, \mathbf C_{\mathcal I} \in \mathbb R^{|\mathcal K| \times d} \} \in \mathbb R^{2|\mathcal K|\times d}$, where $d$ and $|\mathcal K|$ are embedding size and number of intents, respectively. Overall, the space complexity of BIGCF is $O(M+N+2|\mathcal K|\times d)$. Due to the small number of intents (in fact that $|\mathcal K|\ll M$ and $|\mathcal K|\ll N$), the additional number of parameters added by BIGCF is negligible compared with traditional ID embedding-based works (e.g., MF \cite{rendle2009bpr} and LightGCN \cite{he2020lightgcn}).

\subsubsection{\textbf{Time Complexity}}
\begin{table}[t]
\centering
    \setlength{\abovecaptionskip}{0.2cm}
    \setlength{\belowcaptionskip}{0.2cm} 
  \caption{Theoretical time complexity comparisons between BIGCF and seven baselines. The time complexity of embedding encoding is the theoretical time required to complete all operations in a batch. OA stands for additional operation.}
  \label{time}
  \begin{tabular}{l|c|c}
    \hline
    \textbf{Model}&\textbf{Time complexity of encoding}&\textbf{AO}\\
    \hline
    \hline
    {NGCF \cite{wang2019neural}}&$O(2(|\mathcal E|+|\mathcal V|)Ld)$&-\\
    {LightGCN \cite{he2020lightgcn}}&$O(2|\mathcal E|Ld)$&-\\
    {DGCF \cite{wang2020disentangled}}&$O(2|\mathcal K||\mathcal E|Ld)$&-\\
    \hline
    {SGL-ED \cite{wu2021self}}&$O(2(1+2\hat{\rho})|\mathcal E|Ld)$&$O(4\hat{\rho}|\mathcal E|)$\\
    {HCCF \cite{xia2022hypergraph}}&$O(2(|\mathcal E|+H|\mathcal V|)Ld)$&-\\
    {LightGCL \cite{cai2023lightgcl}}&$O(2(|\mathcal E|+q|\mathcal V|)Ld)$&$O(q|\mathcal E|)$\\
    {DCCF \cite{ren2023disentangled}}&$O(2(|E|+|\mathcal K||\mathcal V|)Ld)$&$O(4|\mathcal E|Ld)$\\
    \hline
    \hline
    {\textbf{BIGCF (Ours)}}&$O(2|\mathcal E|Ld +|\mathcal K||\mathcal V|d)$&-\\
    \hline
  \end{tabular}
\end{table}

The time complexity of BIGCF mainly comes from the graph structure encoding and the intent modeling. Specifically, since BIGCF uses LightGCN \cite{he2020lightgcn} as the basic graph encoder, the time complexity of the process is $O(2|\mathcal E|Ld)$, where $|\mathcal E|$ is the number of edges on the interaction graph $\mathcal G$, and the time complexity of the intent modeling part is $O(|\mathcal K||\mathcal V|d)$, where $|\mathcal V| = M+N$ is the number of nodes of the interaction graph $\mathcal G$. Therefore, the total time complexity of BIGCF during training is $O(2|\mathcal E|Ld +|\mathcal K|\mathcal |V|d)$. Table \ref{time} presents the horizontal comparisons of the time complexity of BIGCF with other same-type methods. $\rho$ is the the edge keep probability of SGL-ED \cite{wu2021self}, $H$ is the number of hyperedges in HCCF \cite{xia2022hypergraph}, and $q$ is the required rank for SVD in LightGCL \cite{cai2023lightgcl}. We have the following findings:
\begin{itemize}[leftmargin=*]
\item[$\bullet$] The time complexity of BIGCF is slightly higher than that of LightGCN due to the modeling process of user and item intents.
\item[$\bullet$] The time complexity of BIGCF has a significant advantage over SSL-based methods because the intent modeling is independent of the interaction graph $\mathcal G$ and the number of layers $L$.
\item[$\bullet$] BIGCF introduces the idea of GCL to achieve contrastive regularization in both interaction and intent spaces. The process does not require any additional operations for graph or feature augmentation to construct multi-views, thus significantly reducing the time complexity of model training (\textit{e.g.}, additional graph augmentation is required for SGL-ED \cite{wu2021self} and DCCF \cite{ren2023disentangled}, and SVD-based preprocessing is required for LightGCL \cite{cai2023lightgcl}).

\end{itemize}

\subsubsection{\textbf{Theoretical Analysis}}
\label{analysis}
In this section, we present the positive promotion of the graph contrastive regularization paradigm for BIGCF. Specifically, the graph contrastive regularization process achieves alignment and uniformity in dual spaces with the help of measuring the mutual information $\mathcal I(\mathbf a, \mathbf b)$ between vectors $\mathbf a$ and $\mathbf b$. In the interaction space, for the regularization process of user $u$ and item $i$ in Eq. \ref{graphReg_1}, the gradient of user $u$ is as follows \cite{wu2021self}:
\begin{equation}
\label{user_gradient}
\frac{\partial \mathcal I(\mathbf e_u, \mathbf e_i)}{\partial \mathbf e_u} =\frac{1}{\tau\left \|\mathbf e_u\right\|}\left \{c(u)+ \sum_{j \in \mathcal B/<u,i>}c(j)  \right \},
\end{equation}
where $c(u)$ and $c(j)$ denote the contribution of user $u$ and negative sample ${j}$ to the gradients, respectively. For a large number of negative samples, the $L_2$ norm of $c(j)$ is defined as follows:
\begin{equation}
\label{user_gradient}
\left \| c(j) \right \|_2 \propto \sqrt{1-\text{cos}^2(\mathbf e_u, \mathbf e_j)} \times \text{exp}(\text{cos}(\mathbf e_u, \mathbf e_j)/\tau)  
\end{equation}
The above equation shows that the gradient optimization of $c(j)$ ultimately depends on the cosine similarity. For hard-negative samples $\{j \in \mathcal B/<u,i>\}$, the closer the similarity to $1.0$ will result in a significant increase for the $L_2$ norm $\left \| c(j) \right \|_2$. The above effects will produce a surprisingly exponential increase when moderated by smaller temperature coefficient $\tau$, resulting in a significant improvement in training efficiency \cite{wang2021understanding}.

Noting that in the second and third terms of both Eqs. \ref{graphReg_1} and \ref{graphReg_2}, the two input parameters are identical (\textit{e.g.,} $\mathcal I(\underline{\mathbf e_u, \mathbf e_u})$), we offer the following derivation:

\begin{align}
\mathcal I(\mathbf e_u,\mathbf e_u) & = -\text{log}\frac{\text{exp}\left (\text{cos}(\mathbf e_u, \mathbf e_u)/\tau\right )}{\text{exp}\left (\text{cos}(\mathbf e_u, \mathbf e_u)/\tau\right )+ {\textstyle \sum_{v \in \mathcal B/u}\text{exp}\left (\text{cos}(\mathbf e_u, \mathbf e_v)/\tau \right ) } } \notag\\
&=\left ( -1/\tau +\text{log}\left (\text{exp}\left (1/\tau\right)+ \textstyle \sum_{v \in \mathcal B/u}\text{exp}\left (\text{cos}(\mathbf e_u, \mathbf e_v)/\tau \right )\right) \right ).
\end{align}
This indicates that $\mathcal I(\mathbf e_u,\mathbf e_u)$ actually measures the uniformity among different nodes in a batch. The above corollary also holds for the graph contrastive regularization of item embedding $\mathbf e_i$ and intent embeddings $\{\mathbf e_{u,\boldsymbol{\sigma}},\mathbf e_{i,\boldsymbol{\sigma}}\}$. With gradient optimization, nodes with greater similarity will be penalized more, thus forcing them to move away from each other and making the feature space uniform:
\begin{itemize}[leftmargin=*]
\item[$\bullet$]For interaction embeddings, a more uniform feature space ensures that cold-start items have more chances to be recommended.
\item[$\bullet$]For intent embeddings, a more uniform feature space will ensure that the redundancy among the intents is reduced and better reflects the user's potential preferences. 
\end{itemize}

Furthermore, given final interaction embeddings ($\mathbf e_u$ and $\mathbf e_v$) and individual intent embeddings ($\mathbf e_{u,\boldsymbol{\sigma}}$ and $\mathbf e_{v,\boldsymbol{\sigma}}$) in a mini-batch, we have the following derivations:
\begin{equation}
\begin{aligned}
\label{cos}
\text{cos}(\mathbf e_u, \mathbf e_v) &= \frac{(\mathbf e_{u,\boldsymbol{\mu}} + \mathbf e_{u,\boldsymbol{\sigma}}\odot \boldsymbol{\epsilon}_u)\top(\mathbf e_{v,\boldsymbol{\mu}} + \mathbf e_{v,\boldsymbol{\sigma}}\odot \boldsymbol{\epsilon}_v))}{\left \|\mathbf e_u  \right \|\cdot \left \| \mathbf e_v \right \|};\\
\text{cos} (\mathbf e_{u,\boldsymbol{\sigma}}, \mathbf e_{v,\boldsymbol{\sigma}})&=\frac{(\sum_{k}\mathbb P(\mathbf C_{\mathcal U}^k|\mathbf e_{u,\boldsymbol{\mu}})\cdot \mathbf C_{\mathcal U}^k)^\top(\sum_{k}\mathbb P(\mathbf C_{\mathcal U}^k|\mathbf e_{v,\boldsymbol{\mu}})\cdot \mathbf C_{\mathcal U}^k)}{\left \|\mathbf e_{u,\boldsymbol{\sigma}} \right \| \cdot \left \| \mathbf e_{v,\boldsymbol{\sigma}} \right \|}.
\end{aligned}
\end{equation}
The above equation demonstrates the GCR process across the dual feature spaces, \textit{i.e.}, the process of calculating $\text{cos}(\mathbf e_u, \mathbf e_v)$ will rely on the intent embeddings $\mathbf e_{u,\boldsymbol{\sigma}}$ and $\mathbf e_{,\boldsymbol{\sigma}}$, while the process of calculating $\text{cos} (\mathbf e_{u,\boldsymbol{\sigma}}, \mathbf e_{,\boldsymbol{\sigma}})$ will take into account the interaction embeddings $\mathbf e_{u,\boldsymbol{\mu}}$ and $\mathbf e_{u,\boldsymbol{\mu}}$. The direct (solid line) or non-direct (dashed line) participation process of various types of embeddings executing the graph contrastive regularization is also presented in Fig. 3.

\begin{table}[t]
\small
\setlength{\abovecaptionskip}{0.1cm}
\setlength{\belowcaptionskip}{0.1cm} 
  \caption{ Statistics of the datasets.}
  \label{dataset}
  \begin{tabular}{l|c|c|c|c}
    \hline
    \textbf{Dataset}&\textbf{\#Users}&\textbf{\#Items}&\textbf{\#Interactions}&\textbf{Sparsity}\\
    \hline
    \hline
    \textbf{Gowalla}&50,821&57,440&1,172,425&99.95\%\\
    \textbf{Amazon-Book}&78,578&77,801&2,240,156&99.96\%\\
    \textbf{Tmall}&47,939&41,390&2,357,450&99.88\%\\
    \hline
  \end{tabular}
\end{table}

\begin{table*}
\setlength{\abovecaptionskip}{0.1cm}
\setlength{\belowcaptionskip}{0.1cm} 
  \caption{Overall performance comparisons on Gowalla, Amazon-Book, and Tmall datasets \textit{w.r.t.} Recall@K (abbreviated as R@K) and NDCG@K (abbreviated as N@K). The model that performs best on each dataset and metric is \textbf{bolded}. ‘Improv.\%’ indicates the relative improvement of BIGCF over the best baseline, and the improvement is significant based on two-tailed paired t-test. }
  \label{performance1}
  \begin{tabular}{l|cccc|cccc|cccc}
    \hline
    &\multicolumn{4}{c|}{Gowalla}&\multicolumn{4}{c|}{Amazon-Book}&\multicolumn{4}{c}{Tmall}\\
	\cline{2-13}		
	&R@20&R@40&N@20&N@40&R@20&R@40&N@20&N@40&R@20&R@40&N@20&N@40\\
    \hline
    \hline
	MF \cite{rendle2009bpr}             &0.1553 &0.2264 &0.0923 &0.1108 &0.0557 &0.0873 &0.0411 &0.0525 &0.0465 &0.0755 &0.0316 &0.0417\\
	Mult-VAE \cite{liang2018variational}&0.1793 &0.2535 &0.1079 &0.1267 &0.0736 &0.1126 &0.0559 &0.0687 &0.0531 &0.0841 &0.0365 &0.0473\\
    CVGA \cite{zhang2023revisiting}     &0.1852 &0.2609 &0.1126 &0.1326 &0.0849 &0.1291 &0.0660 &0.0807 &0.0626 &0.0977 &0.0438 &0.0560\\
    \hline
	NGCF \cite{wang2019neural}          &0.1676 &0.2407 &0.0977 &0.1168 &0.0597 &0.0934 &0.0421 &0.0541 &0.0499 &0.0814 &0.0339 &0.0449\\
	LightGCN \cite{he2020lightgcn}      &0.1799 &0.2577 &0.1053 &0.1255 &0.0732 &0.1148 &0.0544 &0.0681 &0.0555 &0.0895 &0.0381 &0.0499\\
    \hline
    DisenGCN \cite{ma2019disentangled}  &0.1379 &0.2003 &0.0798 &0.0961 &0.0481 &0.0776 &0.0353 &0.0451 &0.0422 &0.0688 &0.0285 &0.0377\\
    DisenHAN \cite{wang2020disenhan}    &0.1437 &0.2079 &0.0829 &0.0997 &0.0542 &0.0865 &0.0407 &0.0513 &0.0416 &0.0682 &0.0283 &0.0376\\
    MacridVAE \cite{ma2019learning}     &0.1643 &0.2353 &0.0987 &0.1176 &0.0730 &0.1120 &0.0555 &0.0686 &0.0605 &0.0950 &0.0422 &0.0542\\
    DGCF \cite{wang2020disentangled}    &0.1784 &0.2515 &0.1069 &0.1259 &0.0688 &0.1073 &0.0513 &0.0640 &0.0544 &0.0867 &0.0372 &0.0484\\
    DICE \cite{zheng2021disentangling}  &0.1721&0.2424&0.1025&0.1209&0.0703&0.1129&0.0514&0.0654&0.0601&0.0973&0.0408&0.0536\\
    DGCL \cite{li2021disentangled}      &0.1793 &0.2483 &0.1067 &0.1247 &0.0677 &0.1057 &0.0506 &0.0631 &0.0526 &0.0845 &0.0359 &0.0469\\
	\hline
	SGL-ED \cite{wu2021self}            &0.1809 &0.2559 &0.1067 &0.1262 &0.0774 &0.1204 &0.0578 &0.0719 &0.0574 &0.0919 &0.0393 &0.0513\\
    HCCF \cite{xia2022hypergraph}       &0.1818 &0.2601 &0.1061 &0.1265 &0.0824 &0.1282 &0.0625 &0.0776 &0.0623 &0.0986 &0.0425 &0.0552\\
    LightGCL \cite{cai2023lightgcl}     &0.1825 &0.2601 &0.1077 &0.1280 &0.0836 &0.1280 &0.0643 &0.0790 &0.0632 &0.0971 &0.0444 &0.0562\\
    DCCF \cite{ren2023disentangled}     &0.1876 &0.2644 &0.1123 &0.1323 &0.0889 &0.1343 &0.0680 &0.0829 &0.0668 &0.1042 &0.0469 &0.0598\\
	\hline
	\textbf{BIGCF (Ours)}&\textbf{0.2086} &\textbf{0.2883} & \textbf{0.1242} & \textbf{0.1450} &\textbf{0.0989} &\textbf{0.1468} &\textbf{0.0761} &\textbf{0.0918} &\textbf{0.0755}&\textbf{0.1167} &\textbf{0.0535}&\textbf{0.0680}\\
    \hline
    \hline
    Improv.\%&11.19\%&9.04\%&10.60\%&9.60\%&11.25\%&9.31\%&11.91\%&10.74\%&13.02\%&12.00\%&14.07\%&13.71\%\\
	$p$-value&$2.0\text{e}^{-10}$&$3.0\text{e}^{-8}$&$4.4\text{e}^{-8}$&$1.5\text{e}^{-7}$&$1.9\text{e}^{-6}$&$2.2\text{e}^{-8}$&$7.9\text{e}^{-6}$&$6.5\text{e}^{-8}$&$8.9\text{e}^{-8}$&$5.7\text{e}^{-8}$&$2.8\text{e}^{-7}$&$3.7\text{e}^{-8}$\\
    \hline
  \end{tabular}
\end{table*}

\section{Experiments}

In this section, we perform experiments on three real-world datasets to validate our proposed BIGCF compared with state-of-the-art recommendation methods.

\subsection{Experimental Settings}
\subsubsection{\textbf{Datasets}}
To validate the effectiveness of BIGCF, we adopt three widely used large-scale recommendation datasets: Gowalla, Amazon-Book, and Tmall \cite{ren2023disentangled}, which are varied in scale, field, and sparsity. Table \ref{dataset} provides statistical information for three datasets. To ensure fairness and consistency, we adopt the same processing method with existing efforts \cite{he2020lightgcn, ren2023disentangled}. Specifically, all explicit feedback is forced to be converted to implicit feedback (\textit{i.e.}, ratings are only 0 and 1). Items that a user has interacted with are considered positive samples, and other items apart from that are seen as negative samples for that user. We measure the performance of all recommendation models via Recall@K and NDCG@K \cite{wang2019neural, zhang2023revisiting}.

% \begin{itemize}[leftmargin=*]
% \item[$\bullet$] \textbf{Gowalla} is a classic check-in dataset that contains information about the locations where users share their check-ins.
% \item[$\bullet$] \textbf{Amazon-Book} is a subset of Amazon-reviews about books that contains information about user ratings for products in the books category.
% \item[$\bullet$] \textbf{Tmall} contains data about user purchasing behavior from the online E-platform Tmall.
% \end{itemize}

\subsubsection{\textbf{Baselines}}
To validate the effectiveness of BIGCF, we choose the following state-of-the-art recommendation methods for comparison experiments:
\begin{itemize}[leftmargin=*]
\item[$\bullet$] \textbf{Only factorization-based method}: MF \cite{rendle2009bpr}.
\item[$\bullet$] \textbf{Generative-based methods}: Mult-VAE \cite{liang2018variational} and CVGA \cite{zhang2023revisiting}.
\item[$\bullet$] \textbf{GCN-based methods}: NGCF \cite{wang2019neural} and LightGCN \cite{he2020lightgcn}.
\item[$\bullet$] \textbf{Intent modeling-based methods}: DisenGCN \cite{ma2019disentangled}, DisenHAN \cite{wang2020disenhan}, MacridVAE \cite{ma2019learning}, DGCF \cite{wang2020disentangled}, DICE \cite{zheng2021disentangling}, and DGCL \cite{li2021disentangled}.
\item[$\bullet$] \textbf{SSL-based methods}: SGL-ED \cite{wu2021self}, HCCF \cite{xia2022hypergraph}, LightGCL \cite{cai2023lightgcl}, and DCCF \cite{ren2023disentangled}.
\end{itemize}

\subsubsection{\textbf{Hyperparameter Settings}}
\label{hyper}
We implement BIGCF in PyTorch\footnote{https://github.com/BlueGhostYi/BIGCF}. For a fair comparison, we use an experimental setup consistent with previous works \cite{ren2023disentangled}. Specifically, the embedding size and batch size of all models are set to 32 and 10240, respectively. the default optimizer is Adam \cite{kingma2014adam}, and initialization is done via the Xavier method \cite{glorot2010understanding}. For all comparison methods, we determine the optimal hyperparameters based on the optimal settings given in the paper as well as the grid search. For BIGCF, the number of GCN layers is set in the range of \{1, \textbf{2}, 3\}, the number of intent $|\mathcal K|$ is \{16, 32, \textbf{64}, 128\}, and we empirically set the temperature coefficients $\kappa$ and $\tau$ to be 1.0 and 0.2, respectively. The weight of graph contrastive regularization $\lambda_1$ is set in the range of \{0.1, 0.2, 0.3, 0.4, 0.5\}, and the weight of $L_2$ regularization $\lambda_2$ is set in the range of \{$1\text{e}^{-3}$, $1\text{e}^{-4}$, $\textbf{1e}^{\textbf{-5}}$, $1\text{e}^{-6}$\}. Bolded positions indicate the default optimal settings.

\begin{figure}[t]
\setlength{\abovecaptionskip}{0.1cm}
\setlength{\belowcaptionskip}{0.1cm} 
\centering
\includegraphics[width=\linewidth]{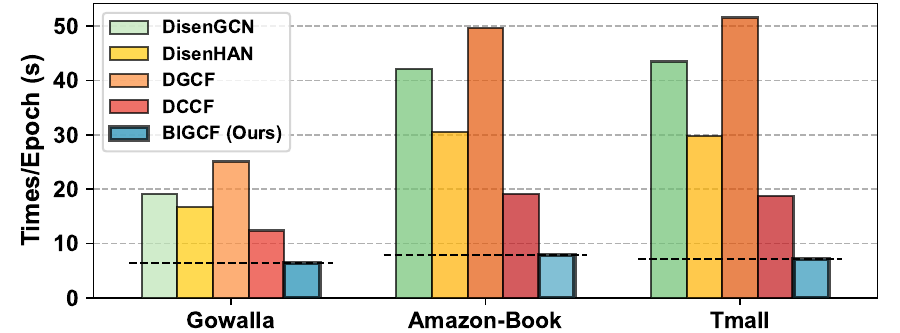}
\caption{Comparison of training time (seconds) per epoch for disentanglement-based methods on three datasets.}
\label{fig_time}
\end{figure}

\subsection{Performance Comparisons}
\subsubsection{\textbf{Overall Comparisons}}
Table \ref{performance1} shows the performance of BIGCF and all baselines, and we have the following observations:
\begin{itemize}[leftmargin=*]
\item[$\bullet$] BIGCF achieves the best performance across all metrics on three datasets. Quantitatively, BIGCF improves by 11.19\%, 11.25\% and 13.02\% over the strongest baseline \textit{w.r.t.} Recall@20 on Gowalla, Amazon-Book, and Tmall datasets, respectively. The above experimental results demonstrate the rationality and generalizability of the proposed BIGCF.
\item[$\bullet$] BIGCF achieves a significant improvement over disentanglement-based approaches (\textit{e.g.}, DGCF \cite{wang2020disentangled} and DCCF \cite{ren2023disentangled}), demonstrating the necessity of modeling intents at a finer granularity.
\item[$\bullet$] BIGCF can achieve better performance than the same type of generative methods (\textit{e.g.}, Mult-VAE \cite{liang2018variational}) and SSL-based methods (\textit{e.g.}, SGL-ED \cite{wu2021self} and LightGCL \cite{cai2023lightgcl}), which demonstrates that the proposed BIGCF can mine user preferences more effectively.
\item[$\bullet$] Fig. \ref{fig_time} presents comparisons of the training time of BIGCF with other disentanglement-based methods, and it can be seen that BIGCF has a significant advantage in training efficiency.
\end{itemize}

\subsubsection{\textbf{Comparisons w.r.t. Data Sparsity}}
\begin{figure}
\setlength{\abovecaptionskip}{0.0cm}
\setlength{\belowcaptionskip}{0.0cm} 
\centering
\subfigure[Gowalla]{\includegraphics[width=1.05in]{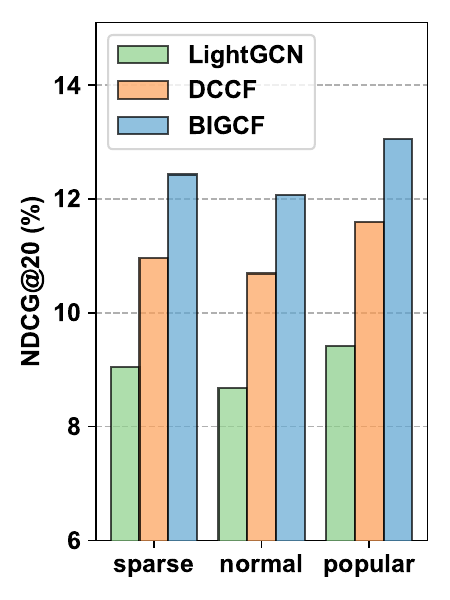}
\label{gowalla_sparsity_case}}
\hfil
\subfigure[Amazon-Book]{\includegraphics[width=1.05in]{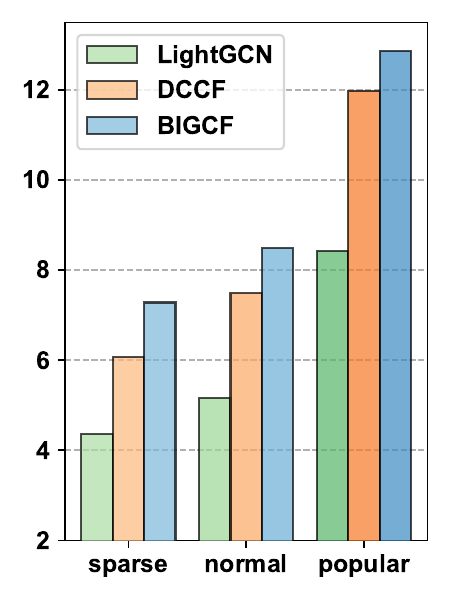}
\label{amazon_sparsity_case}}
\hfil
\subfigure[Tmall]{\includegraphics[width=1.05in]{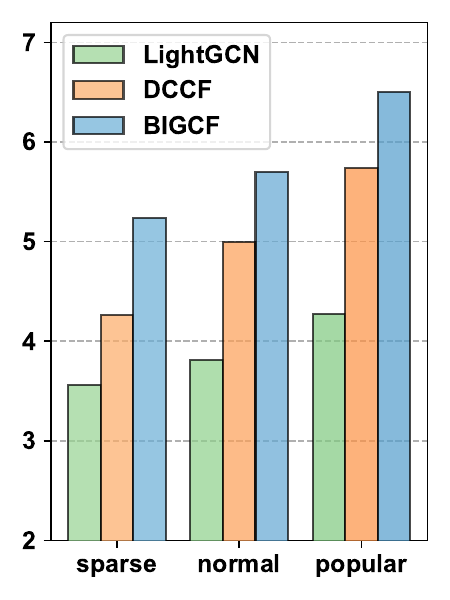}
\label{tmall_sparsity_case}}

\caption{Sparsity tests with three interaction sparsity levels (sparse, normal, and popular) on (a) Gowalla, (b) Amazon-Book, and (c) Tmall datasets \textit{w.r.t.} NDCG@20.  }
\label{fig_sparsity}
\end{figure}

To verify whether proposed BIGCF can effectively deal with the data sparsity issue, we divide the testing set into three subsets based on the interaction scale, denoting sparse users, common users, and popular users, respectively. The experimental results are shown in Fig. \ref{fig_sparsity}. It is clearly evident that BIGCF achieves improvements on all testing subsets of all datasets. Despite the performance degradation of all methods on sparse groups, BIGCF still shows advantages and gradually increases the gap with the performance of DCCF, which indicates that BIGCF has better sparsity resistance. We attribute this to the proposed bilateral intent modeling as well as to the graph contrastive regularization in dual spaces. On the one hand, the finer-grained intents help to deeply understand users' real preferences and prevent the recommender system from over-recommending popular items; on the other hand, the graph contrastive regularization further constrains the distribution of the node representations within the feature space as a way of preventing nodes from over-aggregating.

\begin{table}[!t]

\setlength{\abovecaptionskip}{0.1cm}
\setlength{\belowcaptionskip}{0.1cm} 
\centering
\small
  \caption{Ablation studies of BIGCF on Gowalla, Amazon-Book, and Tmall datasets \textit{w.r.t.} Recall@20 and NDCG@20.}
  \label{tab:ablation}
  \begin{tabular}{l|cc|cc|cc}
    \hline
        &\multicolumn{2}{c|}{Gowalla}&
    \multicolumn{2}{c|}{Amazon-Book}&\multicolumn{2}{c}{Tmall}\\
	\cline{2-7}	
    &R@20&N@20&R@20&N@20&R@20&N@20\\
    \hline
    \hline
    w/o GCR&0.1922&0.1126&0.0783&0.0584&0.0578&0.0395\\
    w/o IR &0.2003&0.1198&0.0957&0.0735&0.0725&0.0514\\
    w/o HNR&0.1931&0.1113&0.0906&0.0676&0.0662&0.0455\\
    w/o BIR&0.1977&0.1182&0.0935&0.0718&0.0727&0.0512\\
    w/o BIGR&0.1991&0.1191&0.0948&0.0720&0.0735&0.0521\\
    w/o PGR&0.1912&0.1131&0.0880&0.0668&0.0680&0.0479\\
    \hline
    \hline
    \textbf{BIGCF}&\textbf{0.2086}&\textbf{0.1242}&\textbf{0.0989}&\textbf{0.0761}&\textbf{0.0755}&\textbf{0.0535}\\

   \hline
\end{tabular}
\end{table}

\subsection{In-depth Studies of BIGCF}
\subsubsection{\textbf{Ablation Studies}}

We construct a series of variants to verify the validity of each module in BIGCF:

\begin{itemize}[leftmargin=*]
\item[$\bullet$] $\text{BIGCF}_{\text{w/o GCR}}$: remove the Graph Contrastive Regularization;
\item[$\bullet$] $\text{BIGCF}_{\text{w/o IR}}$: remove the Interaction Regularization;
\item[$\bullet$] $\text{BIGCF}_{\text{w/o HNR}}$: remove the Homogeneous Node Regularization;
\item[$\bullet$] $\text{BIGCF}_{\text{w/o BIR}}$: remove the Bilateral Intent Regularization;
\item[$\bullet$] $\text{BIGCF}_{\text{w/o BIGR}}$: remove the Bilateral Intent-guided Graph Reconstruction, and use $\mathbf e_{u,\boldsymbol{\mu}}$ and $\mathbf e_{i,\boldsymbol{\mu}}$ directly for downstream tasks;

\item[$\bullet$] $\text{BIGCF}_{\text{w/o PGR}}$: remove the probability-based graph reconstruction (Eq. \ref{final_gcn}) and construct $\mathbf e_u$ and $\mathbf e_i$ by average pooling.
\end{itemize}

The experimental results for all variants with BIGCF on three datasets are shown in Table \ref{tab:ablation} and we have the following findings:

The performance of BIGCF is significantly degraded after removing the GCR module, and it is also degraded after removing the IR, HNR, and BIR modules, respectively. The experimental results demonstrate the necessity and effectiveness of graph contrastive regularization, which can regulate the uniformity of the feature spaces in an augmentation-free and self-supervised manner to prevent the node representations from converging to consistency.

Focusing on intent modeling: (1) firstly, removing the GCR module, $\text{BIGCF}_{\text{w/o GCR}}$ still outperforms LightGCN; (2) secondly, removing the BIR module decreases the performance of BIGCF; (3) lastly, removing the bilateral intent modeling, the performance of $\text{BIGCF}_{\text{w/o BIGR}}$ likewise suffers a certain degree of degradation. The ablation experiments with the above three variants validate the necessity of the proposed intent modeling. Finally, the removal of probabilistic modeling and reparameterization process results in a significant performance degradation of $\text{BIGCF}_{\text{w/o PGR}}$, which demonstrates the effectiveness of generative training strategy.

\begin{figure}
\setlength{\abovecaptionskip}{0.0cm}
\setlength{\belowcaptionskip}{0.0cm} 
\centering
\subfigure[$|\mathcal K|$]{\includegraphics[width=1.62in]{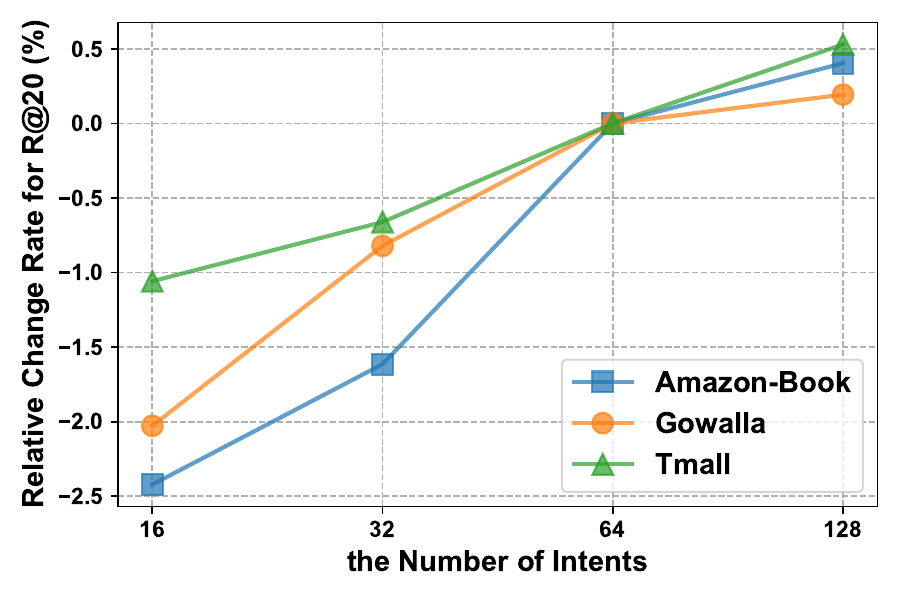}
\label{intent_case}}
\hfil
\subfigure[$\lambda_1$]{\includegraphics[width=1.62in]{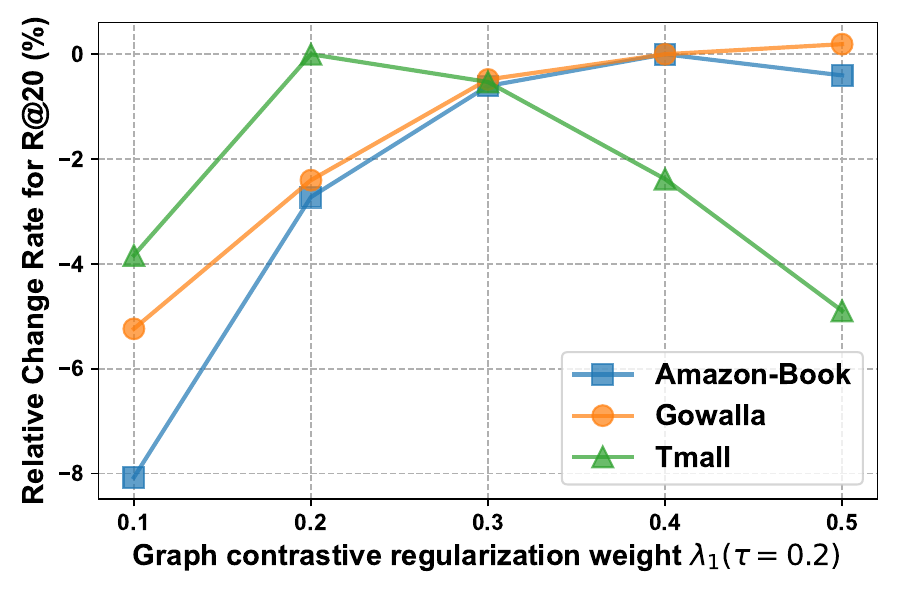}
\label{weight_case}}

\caption{Hyperparameter Sensitivities for (a) the number of commonality intents $|\mathcal K|$ and (b) graph contrastive regularization weight $\lambda_1$ \textit{w.r.t.} Recall@20 on three datasets.}
\label{fig_parameter}
\end{figure}

\begin{table}[!t]

\setlength{\abovecaptionskip}{0.1cm}
\setlength{\belowcaptionskip}{0.1cm} 
\centering
\small
  \caption{Impact on the number of graph convolution layers for BIGCF on Gowalla, Amazon-Book, and Tmall datasets \textit{w.r.t.} Recall@20 and NDCG@20.}
  \label{tab:layer}
  \begin{tabular}{l|cc|cc|cc}
    \hline
        &\multicolumn{2}{c|}{Gowalla}&
    \multicolumn{2}{c|}{Amazon-Book}&\multicolumn{2}{c}{Tmall}\\
	\cline{2-7}	
    &R@20&N@20&R@20&N@20&R@20&N@20\\
    \hline
    \hline
BIGCF-1&0.2052&0.1221&0.0951&0.0736&0.0732&0.0521\\
    \hline
BIGCF-2&\textbf{0.2086}&\textbf{0.1242}&\textbf{0.0989}&\textbf{0.0761}&\textbf{0.0755}&\textbf{0.0535}\\
    \hline
BIGCF-3&0.2066&0.1230&0.0976&0.0750&0.0748&0.0530\\

   \hline
\end{tabular}
\end{table}

\subsubsection{\textbf{Hyperparameter Sensitivities.}}
In this section, we present the effect of various hyperparameters on the performance of BIGCF, as shown in Fig. \ref{fig_parameter} and Table \ref{tab:layer}.

The experimental results in Table \ref{tab:layer} show that BIGCF can effectively fuse graph structure information into the modeling process of user preferences with two GCN layers. However, stacking more GCN layers may introduce unnecessary noise information, which will negatively affect the performance of BIGCF.

As shown in Fig. \ref{fig_parameter}, the performance of BIGCF further improves as the number of intents increases, which proves the effectiveness of intent modeling. When the number increases further, the performance improvement trend of BIGCF starts to slow down, which we argue could be because the excessive number of intents makes the boundaries among intents start to blur, which interferes with the modeling of user preferences. Increasing the strength of the GCR can further improve the performance, which demonstrates the necessity of constraining the uniformity of the feature space. It should be noted that there is a trade-off between alignment and uniformity, and that excessive graph contrastive regularization may cause troubles with the main recommendation task.

\begin{figure}
\setlength{\abovecaptionskip}{0.0cm}
\setlength{\belowcaptionskip}{0.0cm} 
\centering
\subfigure[Intent scores for user $u_{2550}$]{\includegraphics[width=1.62in]{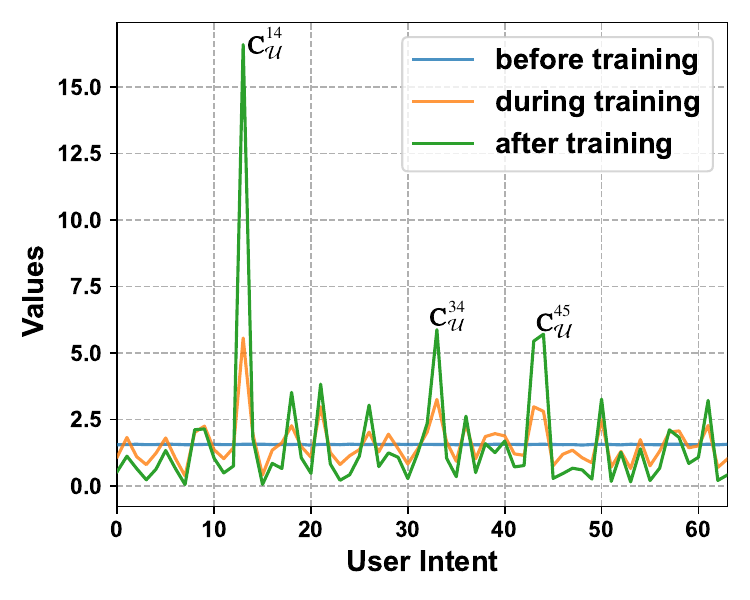}
\label{fig_case_study_1}}
\hfil
\subfigure[Mean and variance of intent scores]{\includegraphics[width=1.62in]{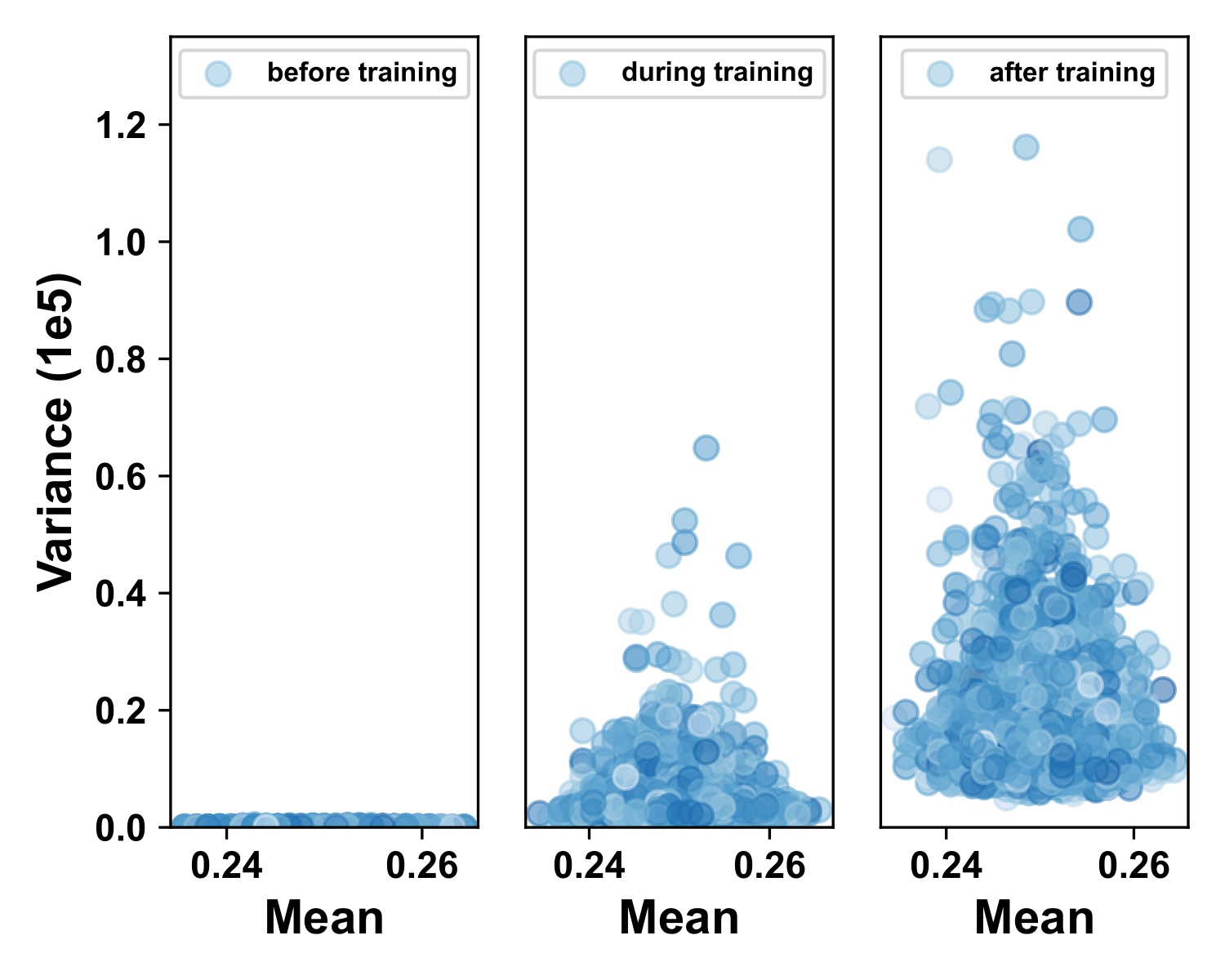}
\label{fig_case_study_2}}
\hfil

\caption{(a) Probability scores of user $u_{2550}$ for collective intent $\mathbf C_{\mathcal U}$. (b) Mean and variance of all user scores for collective intent before, during, and after training.}
\label{fig_case_study}
\end{figure}

\subsubsection{\textbf{Case Study.}} Finally, we evaluate the validity of individual and collective intents through case studies. Specifically, we randomly select a user $u_{2550}$ on Amazon-Book dataset and visualize his correlation scores (computed by Eq. \ref{softmax_mu}) for collective intent (Fig. \ref{fig_case_study_1}). When not yet trained, all intents provide the same contribution, which cannot distinguish the user's true preferences. And as the training proceeds, some aspects that the user does not care about are below the initial value (blue line), while the aspects that he cares about provide larger correlation scores (\textit{e.g.}, $\mathbf C_{\mathcal U}^{14}$, $\mathbf C_{\mathcal U}^{34}$, and $\mathbf C_{\mathcal U}^{45}$). It is worth noting that the user  behavior is not entirely governed by the collective intent $\mathbf C_{\mathcal U}^{14}$, but is also more or less influenced by other intents (\textit{e.g.}, $\mathbf C_{\mathcal U}^{34}$ and $\mathbf C_{\mathcal U}^{45}$). And focusing on overall behavior, we randomly select 5,000 users and show their means and variances of their correlation scores for collective intent before, during, and after training (Fig. \ref{fig_case_study_2}). It can be seen that after training, most users demonstrate greater variances for collective intent, which indicates that each user focuses on different aspects of multiple intents (\textit{e.g.}, user $u_{2550}$ in Fig. \ref{fig_case_study_1}), and BIGCF can effectively utilize implicit feedback data to capture the variability among users and reflect user preferences more accurately.

\section{RELATED WORK}
\textbf{Only ID-based Recommendation.}
Implicit feedback has gradually become a mainstream research component of recommender systems due to its easy accessibility \cite{rendle2009bpr}. With this background, earlier works only map user or item ID to a single embedding and subsequently go through inner product \cite{rendle2009bpr} or neural networks \cite{he2017neural, he2017neural2} to predict user preferences. Many novel designs have also emerged immediately afterward, such as attention \cite{chen2017attentive}, autoencoders \cite{sedhain2015autorec}, and generative models \cite{liang2018variational, zhang2023revisiting}. With the rise of graph neural networks \cite{wu2020comprehensive}, there has been a similar trend of studies in the recommendation field. For example, NGCF \cite{wang2019neural} and LightGCN \cite{he2020lightgcn} use GCNs \cite{kipf2016semi} to model high-quality embedding representations of users and items, and these research results have subsequently spread to other subfields, such as social network \cite{wu2019neural} and knowledge graph-based recommender systems \cite{wang2021learning}. And with the introduction of contrastive learning \cite{chen2020simple}, the research on recommender systems based on graph structures has taken a new step forward. For example, SGL \cite{wu2021self} constructs different views of nodes by random masking, and this structural augmentation can expand the number of samples, thus effectively alleviating the data sparsity problem. Subsequent studies are based on how to simplify the design of SGL, such as replacing structural augmentation with feature augmentation \cite{lin2022improving, cai2023lightgcl}, or designing self-learning augmentation paradigms \cite{xia2022hypergraph}. However, these designs only consider representation modeling and fail to take into account the intents to select items at a fine-grained level, which prevents ID-based methods alone from accurately capturing the user's true preferences.

\textbf{Disentanglement-based Recommendation.}
In general, disent-anglement-based approaches are typically grounded in modeling user-item interactions by projecting them onto different feature spaces \cite{ma2019learning,chen2021curriculum}. For example, MacridVAE \cite{ma2019disentangled} encodes multiple user intents with the variational autoencoders \cite{liang2018variational}. DGCF \cite{wang2020disentangled} and DisenHAN \cite{wang2020disenhan} learn disentangled user representations with graph neural networks and graph attention networks, respectively. DICE \cite{zheng2021disentangling} learns two disentangled causal embeddings for users and items respectively. KGIN \cite{wang2021learning} proposes the notion of shared intents and captures user's path-based intents by introducing item-side knowledge graph. Some novel efforts attempt to integrate contrastive learning into intent modeling, such as ICLRec \cite{chen2022intent}, DiRec \cite{wu2023dual}, and DCCF \cite{ren2023disentangled}.
However, these efforts do not define user or item intents at a fine-grained level, and data augmentation applied to contrastive learning will inevitably make model training more difficult. The proposed BIGCF is contrary to the design philosophy of pioneering works. We present the notion of individuality and collectivity of intents from a causal perspective, and perform intent modeling with a graph generation strategy. In addition, we propose graph contrastive regularization as a simple yet efficient constraint in dual space that does not depend on any type of data augmentation.

\section{Conclusion}
In this paper, we revisited user-item interactions at a fine-grained level and subdivided the motivation for interactions into collective and individual intents with a causal perspective. Based on this, we proposed a novel end-to-end recommendation framework called  Bilateral Intent-guided Graph Collaborative Filtering (BIGCF). The core of BIGCF lies in modeling the bilateral intents of users and items through a graph generation strategy, and these intents are used to guide the interaction graph reconstruction. In addition, we further proposed graph contrastive regularization applicable to both interaction and intent spaces to optimize uniformity among nodes. Finally, we conducted extensive experiments on three real-world datasets and verified the effectiveness of BIGCF.
\begin{acks}
This work is supported by the National Natural Science Foundation of China (No.62272001, No.62206002), the Anhui
Provincial Natural Science Foundation (2208085QF195), and the Hefei Key Common Technology Project (GJ2022GX15).
\end{acks}

\bibliographystyle{ACM-Reference-Format}
\bibliography{sample-base}


\end{document}